\documentclass{amsart}%[article]{iopart}
\usepackage[dvips]{color}
\usepackage{lmodern}
\usepackage[T1]{fontenc}
\usepackage{stmaryrd}
\usepackage{amsmath}%,amssymb}
\usepackage[foot]{amsaddr}
\usepackage{braket}
\definecolor{ddgreen}{rgb}{0.2,0.9,0.0}
\definecolor{turco}{rgb}{0,.5,.5}
\definecolor{reddish}{rgb}{.8,.15,.2}
\definecolor{barney}{rgb}{.6,0.12,0.9}
\definecolor{magic}{cmyk}{0,.7,0,.5}
\definecolor{blueish}{rgb}{.1,.1,.8}
\definecolor{greenish}{rgb}{.2,.4,.2}

\begin{document}

\title[Weyl anomaly of 4D CHS]
{On the Weyl anomaly of 4D Conformal Higher Spins: a holographic approach}
\author{ S Acevedo$^{\$}$, R Aros$^{*}$, F Bugini$^{\S}$ and D E Diaz$^{\dag}$}
\address{${\$}$,${*}$ Departamento de Ciencias Fisicas, Universidad Andres Bello, Sazie 2212, Santiago, Chile}
\address{${\S}$ Departamento de Fisica, Universidad de Concepcion, Casilla 160-C, Concepcion, Chile}
\address{${\dag}$ Departamento de Ciencias Fisicas, Universidad Andres Bello, Autopista Concepcion-Talcahuano 7100, Talcahuano, Chile}
\email{ ${\$}$ sebasti.acevedo@uandresbello.edu,  ${*} $ raros@unab.cl}
\email{ \hspace*{2.6cm}${\S}$ fbugini@udec.cl, ${\dag}$ danilodiaz@unab.cl}

\begin{abstract}
We present a first attempt to derive the full (type-A and type-B) Weyl anomaly of four dimensional conformal higher spin (CHS) fields in a holographic way. We obtain the type-A and type-B Weyl anomaly coefficients for the whole family of 4D CHS fields from the one-loop effective action for massless higher spin (MHS) Fronsdal fields evaluated on a 5D bulk Poincar\'e-Einstein metric with an Einstein metric on its conformal boundary. To gain access to the type-B anomaly coefficient we assume, for practical reasons, a Lichnerowicz-type coupling of the bulk Fronsdal fields with the bulk background Weyl tensor.  Remarkably enough, our holographic findings under this simplifying assumption are certainly not unknown: they match the results previously found on the boundary counterpart under the assumption of factorization of the CHS higher-derivative kinetic operator into Laplacians of ``partially massless'' higher spins on Einstein backgrounds.
\end{abstract}
%\pacs{}

%\submitto{}

\maketitle
\qquad\\

%------------------------------------------------------------------------
\section{Introduction}
\qquad\\

Conformal higher spin (CHS) fields are generalizations of the more familiar 4D Maxwell photon (s=1) and Weyl graviton (s=2) to higher rank $s>2$ totally symmetric tensors. The free field action was introduced by Fradkin and Tseytlin~\cite{Fradkin:1985am} more than thirty year ago; subsequently, cubic interactions were considered in~\cite{Fradkin:1989md} and a fully interacting theory was proposed~\cite{Segal:2002gd} in any dimension $d>2$.
Contrary to massless higher spin (MHS) fields, which have ordinary two-derivative kinetic terms and require (anti-) de Sitter space background~\cite{Fradkin:1987ks,Vasiliev:1990en,Vasiliev:2003ev}, CHS fields exhibit higher-derivative kinetic operators, maximal spin $s$ gauge symmetries, and exist around conformally flat backgrounds, a  conformal class that encompasses Minkowski, de Sitter (dS) and anti-de Sitter (AdS) spacetimes on equal footing.

CHS fields also arise as induced fields in the context of the so-called vectorial AdS/CFT dualities~\cite{Klebanov:2002ja,Sezgin:2002rt,Sezgin:2003pt}, where the elementary CFT fields are N-vectors instead of N-by-N-matrices. There are two ``kinematic'' ways to the CHS action functional at {\em tree level}~\cite{Segal:2002gd,Tseytlin:2002gz,Bekaert:2010ky}. The CHS $h_s$ first show up on the boundary as external sources (or shadow fields) minimally coupled to higher spin singlet bilinear conserved currents $J_s$. After functionally integrating out the vector fields, the UV log-divergent term of the resulting gaussian integration  provides a conformally invariant local action functional for the external sources $h_s$ in even dimensions.
Alternatively, the on-shell action of the bulk dual MHS $\hat{h}_s$ with prescribed boundary asymptotics given by the $h_s$ external currents, i.e. solution of the Dirichlet problem at infinity, contains as well a conformally invariant local action functional for the CHS $h_s$ in the IR log-divergent term in odd-dimensional asymptotically AdS bulk.

In addition, there is also a ``kinematic'' relation at the {\em one-loop} quantum level in the bulk theory and subleading in the large-N expansion of the boundary theory: a {\em holographic formula} relating both $O(1)$-corrected partition functions
\begin{equation}
\frac{Z^{^{(-)}}}{Z^{^{(+)}}}\bigg{|}_{bulk}=Z\bigg{|}_{bndry}
\end{equation}
The bulk side involves the ratio of the functional determinants of the kinetic operator of the bulk field computed with standard and alternate boundary conditions, and the boundary side contains the functional determinant of the kinetic operator of the induced field.
The holographic formula can be thought of as an outgrowth of the IR/UV connection~\cite{Susskind:1998dq} and was reached via a rather circuitous route within AdS/CFT correspondence~\cite{Malda,GKP98,Wit98}, harken back to a class of RG flows triggered by double-trace deformations of the CFT~\cite{Gubser:2002zh,Gubser:2002vv,Hartman:2006dy}.
Ten years after full matching for a massive scalar in Euclidean AdS bulk was shown~\cite{Diaz:2007an}, there are plenty of extensions by now: an incomplete list includes fields with nonzero spin (Dirac, MHS, etc.) and quotients of AdS space (thermal AdS, BTZ, singular AdS, etc.)~\cite{Diaz:2008hy}-\cite{Beccaria:2015uta}~\footnote{In the case of scalars in an
even-dimensional bulk that are duals to boundary GJMS operators, an equivalent result can be found in the conformal geometry literature as well~\cite{Guillarmou05}. The holographic formula, being essentially kinematical, is likely to belong to a broader class of bulk-boundary correspondence as claimed in~\cite{Barvinsky:2005ms,Barvinsky:2014kta,Barvinsky:2017qvf}.}.

Our present interest concerns the holographic formula for MHS (Fronsdal)  fields in the AdS bulk~\footnote{In what follows Euclidean AdS or hyperbolic space H will be meant whenever we write AdS.} and CHS fields on the boundary, as first considered in~\cite{Giombi:2013yva}
\begin{equation}
\frac{Z^{^{(-)}}_{_{\text{MHS}}}}{Z^{^{(+)}}_{_{\text{MHS}}}}\bigg{|}_{_{AdS_5}}=Z_{_{\text{CHS}}}\bigg{|}_{_{S^4}}
\end{equation}
In particular, the type-A Weyl anomaly~\cite{CapperDuff,Deser:1993yx,Boulanger:2007ab} for 4D CHS theory was efficiently obtained in a holographic way from the bulk side of the formula. In AdS the volume factorized and the holographic anomaly was simply given by the volume anomaly~\cite{HS98}-\cite{FG02}  times the one-loop effective Lagrangian for MHS fields, that was already known~\cite{Camporesi:1993mz,Camporesi:1994ga}, together with the contribution from the gauge fixing ghosts. This novel result for the type-A Weyl anomaly follows equivalently from the general holographic prescription of~\cite{ISTY99}. In terms of the number of dynamical degrees of freedom of the CHS gauge field
$\nu_s=s(s+1)$, the compact answer for the Weyl anomaly a-coefficient is a cubic polynomial
\begin{equation}
a_s=\frac{\nu_s^2\,(14\nu_s+3)}{720}
\end{equation}\\
This very same polynomial was reproduced shortly after by a heat kernel computation on the boundary $S^4$~\cite{Tseytlin:2013jya} by exploiting the factorization of the CHS higher-derivative kinetic operator into products of ordinary Laplacians (see also~\cite{Joung:2012qy,Metsaev:2014iwa} for earlier work on this factorization). The relevant heat coefficients $b_4$ for each of the individual (``partially massless'') factors were added up and a remarkable agreement with the holographic prediction was achieved. However, by further assuming factorization of the CHS higher-derivative kinetic operator on Ricci flat backgrounds, motivated by the well known example of the Weyl graviton, the boundary computation also led to the following value for the (shifted) type-B Weyl anomaly coefficient
\begin{equation}
c_s-a_s=\frac{\nu_s\,(15\nu_s^2 - 45\nu_s + 4)}{720}
\end{equation}
with no holographic counterpart whatsoever. In fact, the type-B Weyl anomaly of the boundary induced fields has remained unaccounted for from the holographic perspective in all previously mentioned instances of the holographic formula~\footnote{One notable exception, that inspired the present work, is the case of GJMS operators where a holographic derivation of the 4D and 6D type-B Weyl anomaly has been recently achieved~\cite{BD-GJMS}.}.

The aim of the present work is to bring bulk and boundary approaches into equal footing by providing, for the first time to our knowledge, a plausible holographic derivation of the type-B Weyl anomaly coefficient for CHS fields.
To get access to the type-B Weyl anomaly one needs to go beyond conformal flatness and we do so by allowing for a bulk Poincar\'e-Einstein (PE) metric with an Einstein metric on the boundary conformal class for which the Fefferman-Graham expansion is exact~\cite{Besse,FG12}
\begin{equation}
 \hat{g}_{_{PE}}=\frac{dx^2+(1-\lambda x^2)^2g_{_E}}{x^2}
\end{equation}
where  $\lambda=\frac{R}{4d(d-1)}$ is a multiple of the (necessarily constant) Ricci scalar $R$ of the ($d$-dimensional) boundary Einstein metric and we have set the AdS radius to one. A few words are in order here. As far as CHS fields are concerned, it is likely that they admit a consistent (gauge-covariant) formulation on Bach-flat backgrounds, see~\cite{Nutma:2014pua,Grigoriev:2016bzl,Beccaria:2017nco} for recent progress in this direction. For the purpose of the holographic formula as stated above, a more modest result would suffice, namely, the CHS action at quadratic level in the CHS field but all order in a background boundary Einstein metric (the Einstein condition implies Bach-flatness). The holographic duality, on the other hand, relates this program to the formulation of consistent propagation of MHS fields in the PE background metric. At quadratic level, it requires as well the coupling of the MHS  fields with the bulk background metric, in particular, with the bulk Weyl tensor.
The coupling at the linearized level should be of the Fradkin-Vasiliev type~\cite{Fradkin:1986qy} involving higher derivatives of the fields (dimensionally) compensated by powers of the cosmological constant.
Another important feature that emerges when departing form conformally flat bulk and boundary backgrounds is the presence of  {\em mixing terms} between different spins~\cite{Grigoriev:2016bzl,Beccaria:2017nco} that are likely to be present on both sides of the holographic formula, so that the equality between one-loop partition functions may require the inclusion of the whole family of higher spin fields~\footnote{We thank A. Tseytlin for information on this issue.}.

In this note, with the above caveats in mind, we bring in three crucial elements that pave the way to a holographic derivation of the Weyl anomaly for 4D CHS fields. The first one is that according to the simple recipe put forward in~\cite{Bugini:2016nvn} we only need to look after the coefficient of the Weyl-square bulk term in the one-loop effective Lagrangian, when conveniently written in terms of a basis of conformally covariant curvature invariants, in order to read off $c_s-a_s$\\
\begin{flalign}
\setlength\fboxsep{0.3cm}
\setlength\fboxrule{0.5pt}
\boxed{
\mathcal{A}_4\,\left\{ \,\hat{1}\,,\, \hat{W}^2\,,\, \hat{W}'^{\,3}\,,\,\hat{W}^3,\hat{\Phi}_5\,,\, \hat{W}^4 ...\right\}\,=\,\left\{ \, \frac{1}{16}\,\mathcal{Q}_4\,,\, W^2\,,\, 0\,,\, 0\,,\,0\,,\, 0\,...\right\}
}
\end{flalign}\\
whereas the Euler density content of the type-A is captured by the Q-curvature~\footnote{Let us briefly remind the essence of the holographic recipe from 5D to 4D. One has to look after terms logarithmic in the IR cutoff $\epsilon$ :\\
(i) the radial integral of the PE volume $$\int_{\epsilon}\frac{dx}{x^5} (1-\lambda x^2)^4\,\Rightarrow \lambda^2 \,\log\frac{1}{\epsilon}$$ gives a contribution to the type-A anomaly captured by the Q-curvature;\\
(ii) the bulk $\hat{W}^2=\frac{x^4}{(1-\lambda x^2)^4}W^2$, its radial integral  $$\int_{\epsilon}\frac{dx}{x} W^2\,\Rightarrow W^2 \,\log\frac{1}{\epsilon}$$ gives a contribution to the shifted type-B;\\
(iii) none of the higher conformally covariant curvature invariants ($\hat{W}'^{\,3}\,,\,\hat{W}^3,\hat{\Phi}_5\,,\, \hat{W}^4,...$) will produce a log-term.
}, pure Ricci and equal to $R^2/24$ on a 4D Einstein manifold,\\
\begin{eqnarray}
(4\pi)^{2}\,\langle T\rangle&=&-4\,a\,{\mathcal Q}_4 \,+\,(c-a)\,W^2
\end{eqnarray}\\
The second important element is the assumption of Lichnerowicz-type coupling of the massless bulk field with the background Weyl curvature, we disregard for now the role of any possible additional structure beyond this minimal coupling and possible mixing terms. Finally, the third aspect to be considered is related to WKB exactness of the heat kernel when evaluated on the PE metric. We conjecture that a Weyl-square term can be extracted from each heat kernel coefficient starting with the third (i.e., $\hat{b}_4$) and after resummation we end up with the same exponential pre factor as in the case of AdS for the pure Ricci part. This resummation can be explicitly checked for the third and fourth heat kernel coefficient for the scalar bulk case leading to the Weyl anomaly for GJMS operators~\cite{BD-GJMS}.

The organization of this paper is as follows: we start in section 2 with the holographic derivation of the Weyl anomaly for spin one. In section 3 we address the case of spin two, corresponding to the 5D Einstein graviton and 4D Weyl graviton. We turn to the higher spins in section 4 by first reproducing the anomaly on an Einstein boundary under the assumption of factorization of the higher-derivative kinetic operator. In section 5 we consider the general case of higher spins from the holographic perspective. We finally conclude in section 6 and some miscellaneous results are collected in the appendix~\ref{app.A}.\\

\qquad\\
%----------------------------------------------------------------------------------------------------------------------------
\section{Holographic Weyl Anomaly for spin one}
\qquad\\

Let us start by spelling out the details of the holographic derivation for the gauge vector. The holographic formula for spin one reads
\begin{equation}
\frac{Z^{(-)}_1}{Z^{(+)}_1}\bigg{|}_{_{PE}}\,=\,Z_{_{\text{Maxwell}}}\bigg{|}_{E}
\end{equation}
with the bulk one-loop effective action given by the ratio of functional determinants for the physical and ghost fields
 \begin{align}
Z_1\bigg{|}_{_{PE}}\,=\,\left[\frac{\det\left\{-\hat{\nabla}_{0}^2\right\}} {\det_{\perp}\left\{-\hat{\nabla}_{1}^{2}-4\right\}}\right]^{1/2}
\end{align}

We compute each ``log-det'' with the aid of the (diagonal) heat kernel. Let us first write down the WKB-exact heat expansion in $AdS_5$~\cite{Camporesi:1993mz,Camporesi:1994ga}

\begin{align}
\mbox{spin zero: \qquad tr}\,e^{\{\hat{\nabla}_{0}^{2}\}t}\bigg{|}_{_{AdS_5}}\,=\,\frac{\left(1+\frac{2}{3}t\right)}{(4\pi t)^{5/2}}~e^{-4t}
\end{align}

\begin{align}
\mbox{spin one: \qquad tr}_{\perp}\,e^{\{\hat{\nabla}_{1}^{2}+4\}t}\bigg{|}_{_{AdS_5}}\,=\, \frac{4\left(1+\frac{8}{3}t\right) }{(4\pi t)^{5/2}}~e^{-t}
\end{align}

With this information one can readily get the type-A, but since the bulk is conformally flat (as well as its boundary) any information on the Weyl tensor structure is washed away. We consider therefore the PE metric with nonvanishing Weyl tensor. All pure-Ricci terms will produce the very same answer as in AdS that we already know, so that the information relevant for the type-B anomaly is contained in the terms involving the bulk Weyl tensor and we only need to keep track on the bulk Weyl-square term, in conformity with the holographic recipe~\cite{Bugini:2016nvn}. The heat coefficients for the bulk spin zero field contain Weyl-square contributions starting with the third term $\hat{b}^{(0)}_4$ and it is universally given by
\begin{align}
\mbox{spin zero: \qquad} \hat{b}^{(0)}_4\,\sim \frac{1}{180}\,\hat{W}^2
\end{align}
The spin one requires more care because the standard heat coefficients are given for unconstrained fields and in the present computation we need transverse spin one fields. The unconstrained heat coefficient $\hat{b}^{(1)}_4$ contains two Weyl-square contributions: the same as the scalar for each degree of freedom (5 in 5D) and a contribution from the curvature $\hat{\Omega}_1$ of the spin connection of the vector field (see appendix~\ref{app.A} for details)
\begin{align}
\mbox{\qquad} \hat{b}^{(1)}_4\,\sim \mbox{tr}_{_V}\{I_{_1}\} \,\frac{1}{180}\,\hat{W}^2 + \frac{1}{12}\, \mbox{tr}_{_V}\{\hat{\Omega}_1^2\} \sim
\frac{5}{180}\,\hat{W}^2 - \frac{1}{12}\,\hat{W}^2 = -\frac{1}{18}\,\hat{W}^2
\end{align}
To get the heat coefficient for the transverse 5D spin one field we need to subtract yet the contribution from a minimal scalar (longitudinal mode) so that we end up with
\begin{align}
\mbox{spin one: \qquad} \hat{b}^{(1,\perp)}_4\,\sim -\frac{11}{180}\,\hat{W}^2
\end{align}
One can check that the next heat coefficient $\hat{b}^{(0)}_6$ contains a pure-Ricci part that is captured by the exponential $e^{-4t}$ so that after this factorization the scalar heat kernel in 5D AdS has effectively two terms. The same happens with the Weyl-square term at the PE metric, the $\hat{b}^{(0)}_6$ contains a Weyl-square term that agrees with the combination or convolution of the one in $\hat{b}^{(0)}_4$ and the exponential $e^{-4t}$; the other Weyl invariants (cubic and the ones with derivatives) are such that they do not contribute to the holographic anomaly as noticed in~\cite{BD-GJMS}.
The same WKB exactness will be assumed to be valid for all higher spin bulk fields. In the present spin one case we obtain therefore for the quotient of physical and ghost determinants at the PE metric the following one-loop effective Lagrangian
\begin{equation}
\int_{0}^{\infty}\frac{dt}{t^{7/2}}\left\{e^{-t}\left[4+\frac{32}{3}t - \frac{11}{180}t^2\,\hat{W}^2 + ...\right]-e^{-4t}\left[1+\frac{2}{3}t + \frac{1}{180}t^2\,\hat{W}^2 + ...\right]\right\}
\end{equation}
The ellipsis stands for higher curvature terms in the Weyl tensor that do not contribute to the 4D holographic Weyl anomaly, such as cubic and quartic contractions of the Weyl tensor~\footnote{We should in fact consider a complete basis of conformally covariant bulk curvature invariants completing the seven independent quartic contractions $\hat{W}^4$. However, none of them contribute to the type-B Weyl anomaly in 4D or 6D, so we do not need their explicit form for our current purposes. It is only in 8D where they have  nontrivial `descendants' that match the basis for type-B Weyl invariants of~\cite{Boulanger:2004zf}.}.
After proper time integration, that result in Gamma function factors, we obtain for the one-loop effective Lagrangian (modulo an overall normalization factor)
\begin{equation}
\mathcal{L}^{(s=1)}_{\text{1-loop}}=
-16\cdot\frac{31}{45}\cdot\hat{1} -\frac{13}{180}\cdot\hat{W}^2+...
\end{equation}
The recipe, according to~\cite{Bugini:2016nvn}, is then to read off from the volume or pure-Ricci part $\hat{1}$ the Q-curvature term and from $\hat{W}^2$, simply $W^2$. We end up with the holographic Weyl anomaly for the Maxwell photon
\begin{equation}
\mathcal{A}_4[\text{Maxwell}]= -4\cdot\frac{31}{180}\cdot{\mathcal Q}_4 -\frac{13}{180}\cdot W^2
\end{equation}
so that the Weyl anomaly coefficients turn out to be given by
\begin{equation}
a_1=\frac{31}{180}  \qquad\mbox{and}\qquad c_1-a_1=-\frac{13}{180}
\end{equation}
in agreement with the long-known values for the Maxwell field~\cite{Capper:1974ed,Brown:1977pq}. In particular, for the shifted type-B:
4D vector $b^{(1)}_4\,\sim -\frac{11}{180}\,W^2$ minus ghosts, contributing as two minimal scalar $ b_4^{(0)}\sim\frac{1}{180}\,W^2$ , gives  $b^{(\text{Maxwell})}_4= b^{(1)}_4 - 2b^{(0)}_4\sim-\frac{13}{180}\,W^2$
as follows from the 4D factorization (cf.~\cite{Tseytlin:2013jya})\\
\begin{equation}
Z_{_{\text{Maxwell}}}\bigg{|}_{_{\text{Ricci-flat}}}\,=\,\left[\frac{\left(\det\left\{-\nabla_{0}^2\right\}\right)^{2}} {\det\left\{-\nabla_{1}^2\right\}}\right]^{1/2}\\
\end{equation}\\

\qquad\\
%----------------------------------------------------------------------------------------------------------------------------
\section{Holographic Weyl Anomaly for Spin-Two}
\qquad\\

Let us now turn to the spin two case. The holographic formula for spin two now reads
\begin{equation}
\frac{Z^{(-)}_2}{Z^{(+)}_2}\bigg{|}_{_{PE}}\,=\,Z_{_{\text{Weyl}}}\bigg{|}_{E}
\end{equation}
with the bulk one-loop effective action given by the ratio of functional determinants for the physical and ghost fields
 \begin{align}
Z_2\bigg{|}_{_{PE}}\,=\,\left[\frac{\det_{\perp}\left\{-\hat{\nabla}_{1}^2+4\right\}} {\det_{\perp T}\left\{-\hat{\nabla}_{2}^{2}-2-2\hat{W}\right\}}\right]^{1/2}
\end{align}

The WKB-exact heat expansion in $AdS_5$~\cite{Camporesi:1993mz,Camporesi:1994ga} is given by

\begin{align}
\mbox{spin one: \qquad tr}_{\perp}\,e^{\{\hat{\nabla}_{1}^{2}-4\}t}\bigg{|}_{_{AdS_5}}\,=\,\frac{4\left(1+\frac{8}{3}t\right)}{(4\pi t)^{5/2}}~e^{-9t}
\end{align}

\begin{align}
\mbox{spin two: \qquad tr}_{\perp T}\,e^{\{\hat{\nabla}_{2}^{2}+2\}t}\bigg{|}_{_{AdS_5}}\,=\, \frac{9\left(1+6t\right) }{(4\pi t)^{5/2}}~e^{-4t}
\end{align}

The Weyl-square content of the transverse vector was already found in the spin one case, we recall
\begin{align}
\mbox{spin one: \qquad} \hat{b}^{(1,\perp)}_4\,\sim -\frac{11}{180}\,\hat{W}^2
\end{align}
The unconstrained heat coefficient $\hat{b}^{(2)}_4$ contains three Weyl-square contributions: the same as the scalar for each degree of freedom (5x6/2=15 in 5D), a contribution from the curvature $\hat{\Omega}_2$ of the spin connection of the tensor field and a contribution from the endomorphism (Lichnerowicz coupling $\hat{E}_2=-2\hat{W}$) (see appendix~\ref{app.A} for details)
\begin{align}
\hat{b}^{(2)}_4\sim \,&\mbox{tr}_{_V}\{I_{_2}\} \,\frac{1}{180}\,\hat{W}^2 + \frac{1}{12}\, \mbox{tr}_{_V}\{\hat{\Omega}_2^2\} + \frac{1}{2}\, \mbox{tr}_{_V}\{\hat{E}_2^2\} \\
\sim\,&
\frac{15}{180}\,\hat{W}^2 - \frac{7}{12}\,\hat{W}^2 + \frac{3}{2}\,\hat{W}^2 = \,\hat{W}^2\nonumber
\end{align}
To get to the transverse traceless component we need to subtract the longitudinal part and the trace part~\footnote{We do not write equality sign to stress that we compute modulo pure Ricci curvature invariants.}, so that
\begin{align}
\hat{b}^{(2,\perp T)}_4\sim \,\hat{b}^{(2)}_4 - \hat{b}^{(1)}_4 - \hat{b}^{(0)}_4 \sim \hat{W}^2 - (-\frac{1}{18}\hat{W}^2)-\frac{1}{180}\hat{W}^2
\end{align}
so that we get for transverse traceless spin two
\begin{align}
\mbox{spin two: \qquad} \hat{b}^{(2,\perp T)}_4\,\sim \frac{21}{20}\,\hat{W}^2
\end{align}

For spin two we obtain then for the quotient of physical and ghost determinants at the PE metric the one-loop effective Lagrangian
\begin{align}
\int_{0}^{\infty}\frac{dt}{t^{7/2}}\left\{e^{-4t}\left[9 + 54t + \frac{21}{20}t^2\,\hat{W}^2 + ...\right]-e^{-9t}\left[ 4+\frac{32}{3}t - \frac{11}{180}t^2\,\hat{W}^2 + ...\right]\right\}
\end{align}
The ellipsis stands again for higher curvature terms in the Weyl tensor that do not contribute to the 4D holographic Weyl anomaly.
After proper time integration, that result in Gamma function factors, we obtain for the one-loop effective Lagrangian (modulo an overall normalization factor)
\begin{equation}
\mathcal{L}^{(s=2)}_{\text{1-loop}}=
-16\cdot\frac{87}{5}\cdot\hat{1} + \frac{137}{60}\cdot\hat{W}^2+...
\end{equation}
The holographic recipe~\cite{Bugini:2016nvn} tells us then what the holographic Weyl anomaly for the Weyl graviton is
\begin{equation}
\mathcal{A}_4[\text{Weyl}]= -4\cdot\frac{87}{20}\cdot{\mathcal Q}_4 + \frac{137}{60}\cdot W^2
\end{equation}
and, correspondingly, the Weyl anomaly coefficients
\begin{equation}
a_2=\frac{87}{20}  \qquad\mbox{and}\qquad c_2-a_2=\frac{137}{60}
\end{equation}
in agreement with the well-known results in Conformal Gravity~\cite{Fradkin:1985am,Fradkin:1981iu,Fradkin:1981jc,Fradkin:1982xc}. In particular, for the shifted type-B:
two 4D spin-two traceless (Lichnerowicz), each  $b^{(2,T)}_4\,\sim \frac{21}{20}\,W^2$, minus ghosts, three vectors $b^{(1}_4\,\sim -\frac{11}{180}\,W^2$, resulting in
$b^{(\text{Weyl})}_4= 2b^{(2,T)}_4 - 3b^{(1)}_4\sim\frac{137}{60}\,W^2$, according to the factorized form in 4D (cf.~\cite{Tseytlin:2013jya})
\\
\begin{equation}
Z_{_{\text{Weyl}}}\bigg{|}_{_{\text{Ricci-flat}}}\,=\,\left[\frac{\left(\det\left\{-\nabla_{1}^2\right\}\right)^{3}}{\left(\det_{_T}\left\{-\nabla_{2}^2-2W\right\}\right)^2}\right]^{1/2}\\
\end{equation}\\

\qquad\\
%----------------------------------------------------------------------------------------------------------------------------
\section{Boundary factorization and type-B Weyl anomaly}
\qquad\\

Before extending the previous holographic computations to higher spins, let us first  re-derive the boundary result for $c_s-a_s$ based on factorization of the CHS kinetic operator on Einstein backgrounds~\cite{Tseytlin:2013jya}.
This will allow us to check the validity of our  general results (the three relevant traces) obtained for the heat kernel coefficient that we collect in appendix~\ref{app.A}.

The heat kernel coefficient $b_4$ for the CHS higher-derivative kinetic operator is not available in general, but it has been obtained in~\cite{Tseytlin:2013jya} under the assumption that the factorization on Einstein manifolds observed for the Weyl graviton extends to the whole tower of conformal higher spins. In particular, on Ricci-flat backgrounds one has the following factorized Ansatz in terms of Lichnerowicz Laplacians acting on traceless symmetric tensors\\
\begin{equation}
Z_{_{\text{CHS}}}\bigg{|}_{_{\text{Ricci-flat}}}\,=\,\left[\frac{\left(\det_{_{T}}\left\{-\nabla_{s-1}^2-(s-1)(s-2)W\right\}\right)^{s+1}} {\left(\det_{_{T}}\left\{-\nabla_{s}^2-s(s-1)W\right\}\right)^s}\right]^{1/2}\\
\end{equation}\\
The contributions of the individual Lichnerowicz Laplacians to the Weyl-square term in the $b_4$ heat coefficient, whose numerical coefficient we denote by $\beta$,  come from three terms that for unconstrained fields are given by\\
\begin{align}
b^{(s)}_4&\sim \frac{1}{180} \,\mbox{tr}_{_V}\{I_{_s}\}\,W^2 + \frac{1}{12}\, \mbox{tr}_{_V}\{\Omega_s^2\} + \frac{1}{2}\, \mbox{tr}_{_V}\{E_s^2\} \\
\nonumber\\
 &\sim \beta_s\,W^2=\frac{1}{180}\,\binom{s+3}{3}\,W^2 - \frac{1}{12}\,\binom{s+4}{5}\,W^2 + \frac{3}{2}\,\binom{s+5}{7}\,W^2 \nonumber\\
\nonumber\\
\nonumber
\end{align}
where we have used the general results collected in appendix~\ref{app.A} for the relevant traces.
To get to the traceless symmetric tensors that enter the factorized Ansatz, the trace part must be subtracted\\
\begin{align}\label{b4t}
\beta_s^{^T}&= \beta_{s}-\beta_{s-2}=\frac{(s+1)^2}{720}\,(3s^4+12s^3-2s^2-28s+4)
\end{align}
and finally, the overall coefficient of the Weyl-square term is obtained, according to the alleged factorized form on Ricci flat 4D background, by adding up the contribution from $s$ traceless spin $s$ and subtracting $s+1$ traceless spin $s-1$\\
\begin{align}
\beta_s^{^{CHS}}\,&=\,s\cdot\beta_s^{^T}-(s+1)\cdot\beta_{s-1}^{^T}\\
\nonumber\\
 &=\frac{s(s+1)}{720}\,(15s^4+30s^3-30s^2-45s+4) \nonumber\\
 \nonumber
\end{align}
Therefore, under the assumption of factorization on 4D Einstein backgrounds, the shifted Type-B anomaly for the CHS fields turns out to be\\
\begin{equation}
c_s-a_s=\frac{\nu_s\,(15\nu_s^2 - 45\nu_s + 4)}{720}
\end{equation}
as  originally found in~\cite{Tseytlin:2013jya}, with $\nu_s=s(s+1)$ denoting the number of dynamical degrees of freedom of the CHS gauge field. This is the key result that we will reproduce in a holographic way in what follows.\\

\qquad\\
%----------------------------------------------------------------------------------------------------------------------------
\section{Holographic Weyl anomaly for higher spins}
\qquad\\

Let us now consider the holographic formula for higher spins
\begin{equation}
\frac{Z^{^{(-)}}_{_{\text{MHS}}}}{Z^{^{(+)}}_{_{\text{MHS}}}}\bigg{|}_{_{PE}}\,=\,Z_{_{\text{CHS}}}\bigg{|}_{E}
\end{equation}
with the bulk one-loop effective action given by the ratio of functional determinants for the physical and ghost fields
 \begin{align}
Z_{_{\text{MHS}}}\bigg{|}_{_{PE}}\,=\,\left[\frac{\det_{\perp T}\left\{-\hat{\nabla}_{s-1}^{2}+(s-1)(s+2) -(s-1)(s-2)\hat{W}\right\}} {\det_{\perp T}\left\{-\hat{\nabla}_{s}^{2}-s+(s-2)(s+2)-s(s-1)\hat{W}\right\}}\right]^{1/2}
\end{align}

The WKB-exact heat expansions in $AdS_5$~\cite{Camporesi:1993mz, Camporesi:1994ga} for the ghost and physical transverse traceless fields  are given by

\begin{align}
\mbox{spin $s-1$: \qquad tr}_{\perp T}\,e^{\{\hat{\nabla}_{s-1}^{2}-(s-1)(s+2)\}t}\bigg{|}_{_{AdS_5}}\,=\,\frac{s^2\left(1+\frac{2}{3}s^2t\right)}{(4\pi t)^{5/2}}~e^{-(s+1)^2t}
\end{align}

\begin{align}
\mbox{spin $s$: \qquad tr}_{\perp T}\,e^{\{\hat{\nabla}_{s}^{2}+s-(s-2)(s+2)\}t}\bigg{|}_{_{AdS_5}}\,=\, \frac{(s+1)^2\left(1+\frac{2}{3}(s+1)^2t\right) }{(4\pi t)^{5/2}}~e^{-s^2t}
\end{align}

We need now to determine the Weyl-square content of the transverse traceless CHS fields. We start with the heat kernel coefficient for the unconstrained symmetric tensor in 5D (see appendix~\ref{app.A} for details) \\

\begin{align}
\hat{b}^{(s)}_4&\sim \frac{1}{180} \,\mbox{tr}_{_V}\{I_{_s}\}\,\hat{W}^2 + \frac{1}{12}\, \mbox{tr}_{_V}\{\hat{\Omega}_s^2\} + \frac{1}{2}\, \mbox{tr}_{_V}\{\hat{E}_s^2\} \\
\nonumber\\
 &\sim\hat{\beta}_{_{s}}\,\hat{W}^2=\frac{1}{180}\,\binom{s+4}{4}\,\hat{W}^2 - \frac{1}{12}\,\binom{s+5}{6}\,\hat{W}^2 + \frac{3}{2}\,\binom{s+6}{8}\,\hat{W}^2 \nonumber\\
\nonumber
\end{align}
To get to the transverse traceless component we again need to subtract the longitudinal part and the trace part, so that\\
\begin{align}
\hat{b}^{^{(s,\perp T)}}_4 \sim  \hat{b}^{(s)}_4 - \hat{b}^{(s-1)}_4 - \hat{b}^{(s-2)}_4 + \hat{b}^{(s-3)}_4
\end{align}
and we get for transverse traceless 5D spin $s$ the same numerical coefficient as for traceless 4D spin $s$ (eqn.~\ref{b4t})~\footnote{In fact, to our surprise, we notice that the equality of the Weyl-square coefficients $\hat{\beta}_{s}^{^{\perp T}}$ and $\beta_{s}^{^{T}}$ for  transverse traceless 5D and traceless 4D spin fields, respectively, holds for all three terms in the heat kernel coefficient separately, namely, the scalar Laplacian ($\mbox{tr}_{_V}\{I_{_s}\}$), the curvature of the spin connection ($\mbox{tr}_{_V}\{\Omega^2\}$) and the endomorphism parts ($\mbox{tr}_{_V}\{E^2\}$).}.\\
\begin{align}
\mbox{spin s: \qquad} \hat{b}^{^{(s,\perp T)}}_4\,&\sim\,\hat{\beta}^{^{\perp T}}_{_s}\,\hat{W}^2 \,= \,\frac{(s+1)^2}{720}\,(3s^4+12s^3-2s^2-28s+4)\,\hat{W}^2\nonumber
\end{align}

For the CHS field we obtain then for the quotient of physical and ghost determinants at the PE metric, under the assumption of WKB-exactness, the following one-loop effective Lagrangian\\
\begin{align}
\int_{0}^{\infty}\frac{dt}{t^{7/2}}\left\{e^{-s^2t}\left[(s+1)^2+\frac{2}{3}(s+1)^4t + \hat{\beta}^{^{\perp T}}_{_{s}} \hat{W}^2 t^2 + ...\right]\right.\\
\nonumber\\
\nonumber
\left.-e^{-(s+1)^2t}\left[s^2+\frac{2}{3}s^4t+\hat{\beta}^{^{\perp T}}_{_{s-1}} \hat{W}^2 t^2+ ...\right]\right\}
\end{align}\\
where again the ellipsis stands for higher curvature terms in the Weyl tensor that do not contribute to the 4D holographic Weyl anomaly.
After proper time integration we obtain for the one-loop effective Lagrangian (modulo an overall normalization factor that can be easily worked out)\\
\begin{align}
\mathcal{L}^{^{(\text{CHS})}}_{\text{1-loop}}=&
-16\cdot\frac{s^2(s+1)^2(14s^2+14s+3)}{180}\cdot\hat{1} + \left[s\cdot\hat{\beta}^{^{\perp T}}_{_{s}}-(s+1)\cdot\hat{\beta}^{^{\perp T}}_{_{s-1}}\right]\cdot\hat{W}^2+...\\
\nonumber\\
\nonumber
=&
-16\cdot\frac{s^2(s+1)^2(14s^2+14s+3)}{180}\cdot\hat{1} \\
\nonumber\\
\nonumber
&+ \frac{s(s+1)(15s^4+30s^3-30s^2-45s+4)}{720}\cdot\hat{W}^2+...
\end{align}
The holographic recipe~\cite{Bugini:2016nvn} tells us then that the holographic Weyl anomaly one reads off is simply
\begin{align}
\mathcal{A}_4[\text{CHS}]=& -4\cdot\frac{s^2(s+1)^2(14s^2+14s+3)}{720}\cdot{\mathcal Q}_4 \\
\nonumber\\
\nonumber
&+\, \frac{s(s+1)(15s^4+30s^3-30s^2-45s+4)}{720}\cdot W^2
\end{align}
and, correspondingly, the Weyl anomaly coefficients for the 4D CHS field (with $\nu_s=s(s+1)$) are given by
\begin{equation}
a_s=\frac{\nu_s^2(14\nu_s +3)}{720}  \qquad\mbox{and}\qquad c_s-a_s=\frac{\nu_s(15\nu_s^2-45\nu_s+4)}{720}
\end{equation}
in agreement with the boundary results~\cite{Tseytlin:2013jya}.

\qquad\\
%----------------------------------------------------------------------------------------------------------------------------
\section{Conclusion}
\qquad\\

In all, we have presented a holographic derivation of the Weyl anomaly for Maxwell photon and Weyl graviton. It is based on the one-loop effective action for the dual fields on the bulk PE background. We have learned that the successful match with standard boundary results relies on the WKB-exactness of the heat kernel for bulk spin 1 transverse vector and spin 2 transverse traceless tensor, a prediction that would be worth to explore further.

The extension to CHS fields seems plausible but less sound, since it relies on two simplifying assumptions. The first one is the Lichnerowicz-type coupling with the bulk Well tensor, while consistency with gauge symmetry demands extra interactions terms that involve higher derivatives and whose role is difficult to assess for a series of reasons. For example, possible ambiguities in higher spin theories regarding field redefinitions, total derivatives, etc. need to be considered carefully and, more technically, the heat kernel approach for non-minimal operators (with higher derivatives) may become impractical. The second assumption is that we did not take into account {\em mixing terms} between different spins~\cite{Grigoriev:2016bzl,Beccaria:2017nco} that are likely to be present on both sides of the holographic formula beyond conformally flat bulk and boundary backgrounds, so that the equality may hold only when one includes the whole family of higher spin fields and not `spinwise'.
There are in fact indirect arguments that lead to a different value for the type-B coefficient~\cite{Beccaria:2017lcz} that seems to lead to the vanishing of the total Weyl anomaly, a feature that is favored by virtue of the one-loop quantum consistency of the full CHS theory. Furthermore, under appropriate regularization, it should also be possible to map the finite part of the one-loop effective actions that enter in the holographic formula beyond conformal flatness and, again, a hidden simplicity due to the large higher-spin gauge symmetry is to be expected.

On the other hand, it might well be possible to establish a dictionary for massive bulk fields, without the extra complication of gauge invariance, and in the end the massless limit may yield the correct result for bulk MHS and, correspondingly, boundary CHS fields (see, in this respect,~\cite{Beccaria:2014xda,Beccaria:2014qea,Beccaria:2015uta}).
In any case, further work will be required to settle the many open questions that remain and to explore possible extension to six dimensions and inclusion of half-integer spins.
\qquad\\
%-----------------------------------------------------------------------------
\section*{Acknowledgement}
\qquad\\

We acknowledge useful conversations and correspondence with J. Bellor\'in, N. Boulanger, H. Dorn, R. Olea, S. Ramgoolam, E. Skvortsov, P. Sundell, A. Torrielli, A. Tseytlin and M. Vasiliev. We are particularly grateful to A. Tseytlin for reading through the manuscript and for helpful comments and corrections.
S.A. is a M.Sc. scholarship holder at UNAB. R.A. acknowledges partial support from grants FONDECYT 1151107 and 1140296, UNAB DI 735-15/R and DPI20140115.
The work of F.B. was partially funded by grant CONICYT-PCHA/Doctorado Nacional/2014-21140283. D.E.D. acknowledges support from project UNAB DI 1271-16/R and is also grateful to the Galileo Galilei Institute for Theoretical Physics (GGI) for the hospitality and INFN for partial support during the stay at the program ``New Developments in AdS3/CFT2 Holography'' and to the Quantum Field and String Theory Group at Humboldt University of Berlin for the kind invitation and the opportunity to present the results reported here.

\qquad\\
\newpage
\appendix
\section{Heat kernel coefficient $b_4$ for unconstrained totally symmetric rank s tensors}\label{app.A}
\qquad\\

Let us consider the Laplacian $-\nabla_{s}^2$ and a matrix-valued potential (endomorphism) $E_s$ acting on unconstrained totally symmetric rank s tensors in $D$ dimensions
\begin{equation}
-\nabla_{s}^2-E_s
\end{equation}
The (diagonal) heat kernel coefficient $b_4$ has then three independent contributions to the Weyl-square term if we asume a Lichnerowicz-type coupling with the background Weyl tensor $E_s=s(s-1)W$:  the same as the scalar for each degree of freedom, a contribution from the curvature $\hat{\Omega}_s$ of the spin connection of the tensor field and a contribution from the endomorphism $E_s$
\begin{align}
b^{(s)}_4&\sim \frac{1}{180} \,\mbox{tr}_{_V}\{I_{_s}\}\,W^2 + \frac{1}{12}\, \mbox{tr}_{_V}\{\Omega_s^2\} + \frac{1}{2}\, \mbox{tr}_{_V}\{E_s^2\} \\
\nonumber
\end{align}
The numerical coefficients above are universal~\cite{Gilkey:1975iq}, the dimensionality $D$ of the space enters only via the traces.
We compute them separately as follows.\\

\paragraph{\bf Dimensionality} The first trace simply counts the number of components of a totally symmetric spin-s field in $D$ dimensions (Young tableau consisting of a single row of length $s$)

\begin{align}
\mbox{tr}_{_V}\{I_{_s}\}\,=\,\binom {D+s-1}{D-1}
\end{align}
The low-spin examples are well known:

\begin{align}
\mbox{spin zero: \qquad tr}_{_V}\{I_{_0}\}\,&=\, 1\\
\nonumber\\
\mbox{spin one: \qquad tr}_{_V}\{I_{_1}\}\,&=\, D\\
\nonumber\\
\mbox{spin two: \qquad tr}_{_V}\{I_{_2}\}\,&=\, \frac{D(D+1)}{2}
\end{align}

\qquad\\
\paragraph{\bf Endomorphism} The index structure of the endomorphism (the piece in the Lichnerowicz coupling containing the Weyl tensor) is as follows\\
\begin{equation}
E_s= s(s-1)W^{\rho_{1}\,\rho_{2}}_{~\nu_{1}~\nu_{2}}\delta_{\rho_{1}}^{(\mu_{1}}\delta_{\rho_{2}}^{\mu_{2}}\delta_{\nu_{3}}^{\mu_{3}}\cdots\delta_{\nu_{s}}^{\mu_{s})}\\
\end{equation}\\
This clearly generalizes the spin two for which one actually has the part involving the Riemann tensor $2R^{(c\,d)}_{~a~b}$ (see e.g.~\cite{Avramidi:2015pqa}) that gives the only Weyl-tensor contribution $2\,W^{(c\,d)}_{~a~b}$

\begin{align}
\mbox{spin two: \qquad} E_2=2\,W^{\rho_{1}\,\rho_{2}}_{~a~b}\delta_{\rho_{1}}^{(c}\delta_{\rho_{2}}^{d)}
\end{align}\\
In taking the trace of the square of the endomorphism, the totally symmetrized product of Kronecker deltas can be compactly and conveniently written in terms of the {\em generalized permanent delta} (gpd) $\Pi$ (see e.g.~\cite{agacy1999generalized})~\footnote{In fact, the trace of the identity endomorphism $I_{_s}=\delta_{\nu_{1}}^{(\mu_{1}}\delta_{\nu_{2}}^{\mu_{2}}\delta_{\nu_{3}}^{\mu_{3}}\cdots\delta_{\nu_{s}}^{\mu_{s})}= \frac{1}{s!}\Pi_{\nu_{1}\nu_{2}\nu_{3}\cdots\nu_{s}}^{\mu_{1}\mu_{2}\mu_{3}\cdots\mu_{s}}$ also follows easily from the properties of the gpd $\mbox{tr}_{_V}\{I_{_s}\}\,=\,\frac{1}{s!}\,\Pi_{\mu_{1}\mu_{2}\mu_{3}\cdots\mu_{s}}^{\mu_{1}\mu_{2}\mu_{3}\cdots\mu_{s}}=\frac{(D+s-1)!}{s!(D-1)!}$ . }

\begin{eqnarray}
\qquad \mbox{tr}_{_V}\{E_s^2) &=& s^{2}(s-1)^{2}W^{\rho_{1}\rho_{2}}_{~\nu_{1}~\nu_{2}}\delta_{\rho_{1}}^{(\mu_{1}}\delta_{\rho_{2}}^{\mu_{2}}\delta_{\nu_{3}}^{\mu_{3}}\cdots\delta_{\nu_{s}}^{\mu_{s})}
W^{\lambda_{1}\lambda_{2}}_{~\mu_{1}~\mu_{2}}\delta_{\lambda_{1}}^{(\nu_{1}}\delta_{\lambda_{2}}^{\nu_{2}}\delta_{\mu_{3}}^{\nu_{3}}\cdots\delta_{\mu_{s}}^{\nu_{s})}\\
 &=& \frac{s^{2}(s-1)^{2}}{(s!)^{2}}W^{\rho_{1}\rho_{2}}_{~\nu_{1}~\nu_{2}}W^{\lambda_{1}\lambda_{2}}_{~\mu_{1}~\mu_{2}}
 \Pi_{\rho_{1}\rho_{2}\nu_{3}\cdots\nu_{s}}^{\mu_{1}\mu_{2}\mu_{3}\cdots\mu_{s}}\Pi_{\lambda_{1}\lambda_{2}\mu_{3}\cdots\mu_{s}}^{\nu_{1}\nu_{2}\nu_{3}\cdots\nu_{s}}\nonumber
\end{eqnarray}\\
We proceed now by contracting the permanents to extract the overall dependence on $s$ and $D$ until we reach the $s=2$ term for which we already know the answer (see e.g.~\cite{Avramidi:2015pqa}, eqn.3.50)

\begin{align}
\mbox{spin two: \qquad} \mbox{tr}_{_V}\{E_2^2)=3\,W^2
\end{align}\\
The two relevant identities~\footnote{The second identity is eqn.(p.5) in~\cite{agacy1999generalized}, whereas the first one is actually an extension of eqn.(p.13) that includes extra `spectator' indices $l$'s and $m$'s. Incidentally, diagrammatic methods for representing linear operators in tensor spaces, their products and traces (see, e.g.,~\cite{Ramgoolam:2016ciq} and references therein) are likely to provide a most elegant derivation of the previous results. We are indebted to S. Ramgoolam for valuable explanations on this possibility.} for the permanents are the following

 \begin{align}
 \Pi_{j_{1}\cdots j_{p} ~j_{p+1}\cdots j_{q}}^{i_{1}\cdots i_{p} ~i_{p+1}\cdots i_{q}}\Pi_{k_{p+1}\cdots k_{q} ~m_{1}\cdots m_{r}}^{j_{p+1}\cdots j_{q} ~l_{1}\cdots l_{r}}=(q-p)!~\Pi_{j_{1}\cdots j_{p}~k_{p+1}\cdots k_{q}~m_{1}\cdots m_{r}}^{i_{1}\cdots i_{p}~i_{p+1}\cdots i_{q}~l_{1}\cdots l_{r}}
 \end{align}

 \begin{align}
 \Pi_{j_{1}\cdots j_{p}~i_{p+1}\cdots i_{q}}^{i_{1}\cdots i_{p}~i_{p+1}\cdots i_{q}}=\frac{(D+q-1)!}{(D+p-1)!}\Pi_{j_{1}\cdots j_{p}}^{i_{1}\cdots i_{p}}
 \end{align}\\
We take $q=s$ and $p=2$ in the first identity above and then $q=s+2$ and $p=4$ in the second to get

 \begin{eqnarray}
 \Pi_{\rho_{1}\rho_{2}\nu_{3}\cdots\nu_{s}}^{\mu_{1}\mu_{2}\mu_{3}\cdots\mu_{s}}\Pi_{\lambda_{1}\lambda_{2}\mu_{3}\cdots\mu_{s}}^{\nu_{1}\nu_{2}\nu_{3}\cdots\nu_{s}} &=& (s-2)!\,\Pi_{\lambda_{1}\lambda_{2}\rho_{1}\rho_{2}\mu_{3}\cdots\mu_{s}}^{\nu_{1}\nu_{2}\mu_{1}\mu_{2}\mu_{3}\cdots\mu_{s}} \label{eq.1}\\
  &=& (s-2)!\frac{(D+s+1)!}{(D+3)!}\Pi_{\rho_{1}\rho_{2}\lambda_{1}\lambda_{2}}^{\mu_{1}\mu_{2}\nu_{1}\nu_{2}} \nonumber
 \end{eqnarray}\\
The remaining gpd is the one for spin two that produced $3\,W^2$, therefore the trace turns out to be

 \begin{align}
\mbox{tr}_{_V}\{E_s^2\}&=\frac{s^{2}(s-1)^{2}(s-2)!(D+s+1)!}{(s!)^{2}(D+3)!}\cdot3\,W^2\\
\nonumber
 &=3\,\binom {D+s+1}{D+3}\,W^2
\end{align}

\qquad\\
\paragraph{\bf Curvature} The curvature of the spin connection for the rank s totally symmetric tensor is given by\\
\begin{equation}
\Omega_s= \frac{1}{2}R_{\mu\nu}^{\;\;\; ab}\,\Sigma_{ab}
\end{equation}\\
with $\Sigma_{ab}$ realizing the Lorentz algebra\\
\begin{align}
[\Sigma_{ab} ,\Sigma_{cd}]=-g_{ac}\Sigma_{bd}+g_{ad}\Sigma_{bc}+g_{bc}\Sigma_{ad}-g_{bd}\Sigma_{ac}
\end{align}\\
The explicit index structure of the curvature, generalizing spins one and two, is given by
\begin{equation}
(\Sigma_{ab})_{c_{1}\cdots c_{s}}^{d_{1}\cdots d_{s}}= 4\, \delta_{[a}^{(d_{1}}g_{b](c_{1}}\delta_{c_{2}}^{d_{2}}\cdots\delta_{c_{s})}^{d_{s})}
\end{equation}\\
We need now to take the trace of the curvature square
\begin{align}
\mbox{tr}_{_V}\{\Omega_s^2\}=\frac{1}{2}R_{\mu\nu}^{~ab}(\Sigma_{ab})_{c_{1}\cdots c_{s}}^{d_{1}\cdots d_{s}}\frac{1}{2}R^{\mu\nu cd}(\Sigma_{cd})_{d_{1}\cdots d_{s}}^{c_{1}\cdots c_{s}}
\end{align}
We first write in terms of gpd's to proceed again by contracting to extract the overall dependence on $s$ and $D$ until we reach now the $s=1$ term for which we also already know the answer (see e.g.~\cite{Avramidi:2015pqa}, eqn.3.12)

\begin{align}
\mbox{spin one: \qquad} \mbox{tr}_{_V}\{\Omega_1^2)\sim -\,W^2
\end{align}\\
Modulo overall factor and anti-symmetrization in the indices $[a\,b]$ and $[c\,d]$, we consider the product that enters in the trace\\
\begin{align}
&g_{b\hat{d}}\,\delta_{a}^{(d_{1}|}\,\delta_{(c_{1}}^{\hat{d}}\delta_{c_{2}}^{|d_{2}}\cdots\delta_{c_{s})}^{d_{s})}\,\,g_{d\hat{c}}\,\delta_{c}^{(c_1|}\delta_{(d_{1}}^{\hat{c}}\delta_{d_{2}}^{|c_{2}}
\cdots\delta_{d_{s})}^{c_{s})}\\
\nonumber\\
%\nonumber
&=g_{b\hat{d}}\,\,\delta_{(c}^{\hat{d}}\delta_{c_{2}}^{(d_{2}}\cdots\delta_{c_{s})}^{d_{s})}\,\,g_{d\hat{c}}\,\delta_{(a}^{\hat{c}}\delta_{d_{2}}^{(c_{2}}
\cdots\delta_{d_{s})}^{c_{s})}\\
\nonumber\\
%\nonumber
&= g_{b\hat{d}}\,\frac{\Pi_{c c_{2}\cdots c_{s}}^{\hat{d} d_{2}\cdots d_{s}}}{(s-1)!} \,\,g_{d\hat{c}}\,\frac{\Pi_{a d_{2}\cdots d_{s}}^{\hat{c} c_{2}\cdots c_{s}}}{(s-1)!}\\
\nonumber\\
%\nonumber
&= g_{b\hat{d}}\,g_{d\hat{c}}\,\frac{1}{(s-1)!}\,\Pi_{a c c_{2}\cdots c_{s}}^{\hat{d}\hat{c} c_{2}\cdots c_{s}}\\
\nonumber\\
%\nonumber
&= g_{b\hat{d}}\,g_{d\hat{c}}\,\Pi_{a c}^{\hat{d}\hat{c}}\,\frac{1}{(s-1)!}\,\frac{(D+s)!}{(D+1)!}
\end{align}\\
Now that  we have extracted the overall dependence on $s$ and $D$, the rest produces the spin one answer. Finally, we obtain the curvature contribution to the Weyl-square term\\
\begin{align}
\mbox{tr}_{_V}\{\Omega_s^2\}\sim\,-\binom {D+s}{D+1}W^2
\end{align}\\
This yields for unconstrained spin two fields\\
\begin{align}
\mbox{spin two: \qquad} \mbox{tr}_{_V}\{\Omega_2^2)\sim -\,(D+2)\,W^2
\end{align}\\
in accordance, for example, with~\cite{Avramidi:2015pqa} (eqn.3.54).

We close by noting that the above combinatorial coefficients comprise several scattered results in the literature.
For example, for Lichnerowicz Laplacians on traverse traceless tensors one has to subtract the trace and the longitudinal contributions as we did in 5D in section 5\\
\begin{align}
\beta^{^{\perp T}}_{_{s}}\,=\,&\quad\frac{1}{180}\,\binom{D+s-1}{D-1}\,-\,\frac{1}{12}\,\binom{D+s}{D+1}\, +\, \frac{3}{2}\,\binom{D+s+1}{D+3}\\
\nonumber\\
\nonumber
 &-\frac{1}{180}\,\binom{D+s-2}{D-1}\,+\,\frac{1}{12}\,\binom{D+s-1}{D+1}\, -\, \frac{3}{2}\,\binom{D+s}{D+3}\\ \nonumber\\
\nonumber
&-\frac{1}{180}\,\binom{D+s-3}{D-1}\,+\,\frac{1}{12}\,\binom{D+s-2}{D+1}\, -\, \frac{3}{2}\,\binom{D+s-1}{D+3}\\ \nonumber\\
\nonumber
&+\frac{1}{180}\,\binom{D+s-4}{D-1}\,-\,\frac{1}{12}\,\binom{D+s-3}{D+1}\, +\, \frac{3}{2}\,\binom{D+s-2}{D+3}%\\
%\nonumber\\
%\nonumber
%&=...
\end{align}\\
Below we report the low spin examples, in agreement with~\cite{Ohta:2016jvw} (appendix A therein),

\begin{align}
\mbox{spin zero: \qquad}& \beta_{_{0}}\,=\, \frac{1}{180}\\
\nonumber\\
\mbox{spin one: \qquad}&\beta^{^{\perp }}_{_{1}}\,=\, \frac{D-16}{180}\\
\nonumber\\
\mbox{spin two: \qquad}&\beta^{^{\perp T}}_{_{2}}\,=\, \frac{D^2-31D+508}{360}
\end{align}

\qquad\\
%------------------------------------------------------------------------------------

\hspace{0.5cm}

\providecommand{\href}[2]{#2}\begingroup\raggedright\endgroup

\end{document}